# Proton and Neutron Induced SEU Cross Section Modeling and Simulation: A Unified Analytical Approach

**Gennady I. Zebrev [1], Nikolay N. Samotaev [1], Rustem G. Useinov [2], Artur M. Galimov [3], Vladimir V. Emeliyanov [2], Artyom A. Sharapov [2], Dmitri A. Kazyakin [1,3], and Alexander S. Rodin [1]**

[1] Micro- and Nanoelectronics Department, National Research Nuclear University MEPhI (Moscow Engineering Physics Institute), Moscow, Russia; gizebrev@mephi.ru (G.Z.); nnsamotaev@mephi.ru (N.S.); asrodin@mephi.ru (A.R.);
[2] Research Institute of Scientific Instruments, Lytkarino, Moscow region, Russia; rust.us@mail.ru (R.U.); vemeliyanov@niipribor.ru (V.E.)
[3] MRI Progress, Moscow, Russia; armgalimov@gmail.com
* Correspondence: nnsamotaev@mephi.ru

**Abstract:** A new physics-based compact model, which makes it possible to simulate in a unified way the neutron and proton of cosmic ray induced SEU cross sections, including effects from nuclear reaction products and from direct ionization by low-energy protons, has been proposed and validated. The proposed approach is analytical and based on explicit analytical relationships and approximations with physics-based fitting parameters. GEANT4 or SRIM numerical calculations can be used as an aid to adjust or refine the phenomenological parameters or functions included in the model taking into account real geometrical configurations and chemical compositions of the devices. In particular, explicit energy dependencies of the soft error cross sections for protons and neutrons over a wide range of nucleon energies were obtained and validated. Main application areas of developed model include space physics, accelerator studies high energy physics and nuclear experiments.

**Keywords:** direct ionization, low-energy protons, modeling, neutrons, nuclear reactions, SEU cross sections, single event upsets

## 1. Introduction

Soft errors (or Single Event Upsets, SEUs), i.e. the reversible changes of digital devices' states caused by one single ionizing particle, are one of the main challenges of modern digital electronics [1]. One of the main sources of such errors on board of space satellites and in avionics is the nuclear interactions of protons and neutrons with atoms of the material of electronic circuits. Soft Error Rate (SER) or cross section prediction of nucleon-induced SEU faces great difficulties because the circuit and the nuclear interaction simulators are poorly compatible technically and conceptually. Besides, aggressive scaling of the microelectronics components leads to decreased immunity of the digital integrated circuits to external transients due to reducing in noise margin due to lowering of supply voltage and element size reduction. In particular, the commercial highly-scaled digital memories become extremely susceptible to the single event effects (SEE). For example, the critical charge $Q_C$, i.e. the minimum charge to cause an upset of the memory cell is reducing to the sub-femtoCoulomb region. Such values of the collected charge (of order $10^3 - 10^4$ carriers) correspond to the mean deposited energy as small as a few keV and average values of critical linear energy transfer (LET) less than 1 MeV-cm$^2$/mg. For example, the circuit critical charge $\varepsilon_C$ can be mapped to the transferred critical energy required to switch the digital element [2].

$$\varepsilon_C = \frac{Q_C}{q}\varepsilon_p = 22.5\left(\frac{Q_C}{fC}\right)\ \text{keV}, \qquad (1)$$

where $\varepsilon_p \cong$ 3.6 eV(Si) is a mean energy of electron-hole creation [2]. It is important to emphasize that $\varepsilon_C$ and $Q_C$ are, as a rule, the purely circuit parameters that characterizes immunity of a typical digital node to transient electrical disturbances of any origin. The probability (cross section) of a nucleon-induced error becomes noticeable when the local energy release from the products of a nuclear reaction and Linear Energy Transfer (LET) of nuclear reaction products exceed a critical value $\Lambda_{\sec} > \Lambda_C \propto \varepsilon_C$, where $\Lambda_{\sec}$ is a typical LET of secondary particles. Since the LET of direct ionization from protons is small, it has long been reasonably assumed that when exposed to protons, a necessary condition for soft error generation is a proton-induced nuclear reaction. Nevertheless, the technologically driven decrease in $\varepsilon_C$ and $Q_C$ leads to the fact that single event upsets in digital circuits can be caused by direct ionization of particles even with a very small LET value, for example, by low-energy protons [3-5].

Another challenge is the neutron-induced soft error rate (SER), arising from the finite value of the neutron flux in atmosphere, or even at the sea level (~ 13 $cm^{-2}$ $h^{-1}$) [6-9]. The physical processes involved in the failure mechanisms from neutrons and low-energy protons are quite complex and strictly speaking require cumbersome computational approaches based on numerical simulations (e.g., GEANT4, TCAD, or MRED) [10-12]. Such systems are often inaccessible, inconvenient to use and require highly skilled users.

In this work, we propose a variant of a compact model for calculation of the SEU cross sections induced by protons, neutrons and direct ionization (including by low energy protons) based on transparent physical models and controlled approximations. We will demonstrate here that even such seemingly heterogeneous problems as modeling of the SEU cross sections caused by secondary products of nuclear reactions and by low-energy proton direct ionization can be readily simulated within the framework of a unified approach. The main motivation of this work is to try to replace the cumbersome and opaque GEANT4 numerical simulation with a simple and transparent analytical model with physically measured and calculated parameters.

## 2. Methodology

### *2.1. A General Framework*

The suggested model in its first is intended to analyze SEU rate in regularly arranged electronic devices such as memory circuits. There are two types of the single event errors, namely, the errors due to direct ionization of primary particles, and errors due to ionization by the secondary products of elastic or non-elastic nuclear interactions formed upon irradiation with the primary particles. Both types are determined ultimately by direct ionization with heavy ions of either primary or secondary origin. The volume in which charge separation and collection occurs is commonly referred to as the sensitive volume [13]. Due to the nonlocality of the ion impact, the concept of a sensitive volume for an individual memory cell is violated, since the entire memory area is a single sensitive volume with an effective thickness $t_{eff}$ [14]. This effective thickness turns out to be quite small (≥10 nm) for ICs with sub-femtoCoulomb $Q_C$. At the same time, the thickness of the region in which nuclear reactions occur, affecting the ionization in the sensitive region, is of the order of the secondary particle range ($L_R$≥1 μm). Conceptually, these are two completely different regions, one of which is determined by circuit factors (simulated with CAD or TCAD), and the other by nuclear physics (GEANT4). The region of actual nuclear reactions will be referred to below as the influencing region (see Fig. 1).

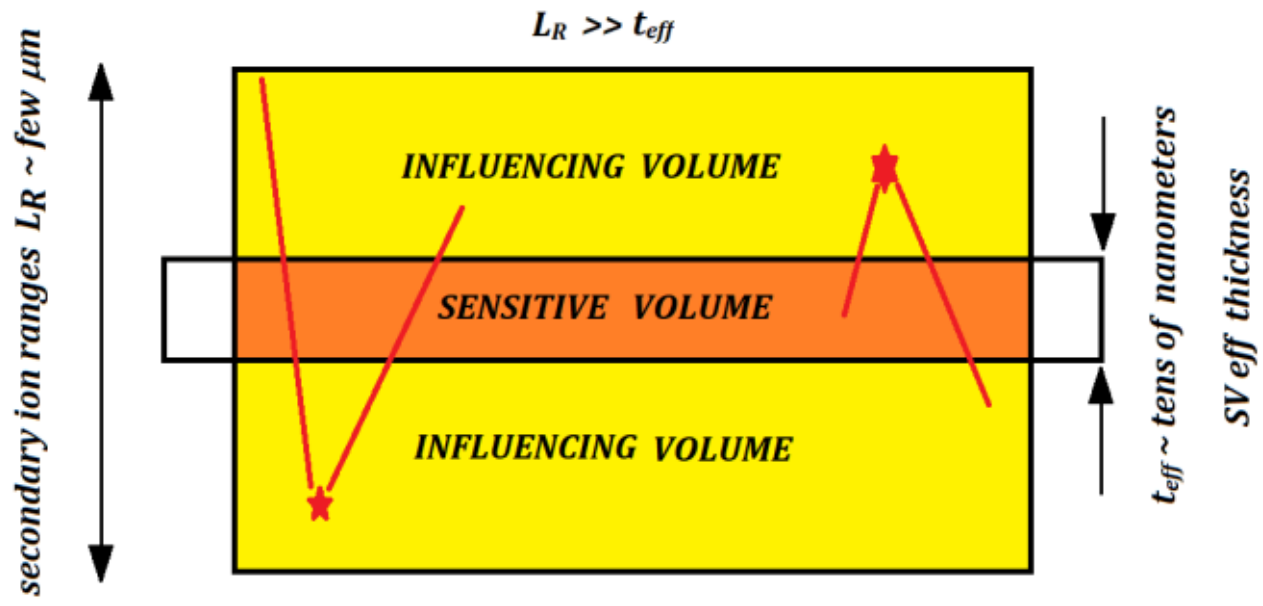

**Figure 1.** Conceptual view of influencing and sensitive volumes.

The influencing volume, where actual nuclear reactions take place, are typically much larger than the sensitive volume where is the charge collected ( $L_R >> t_{eff}$ ). This allows us to represent the sensitive region as being affected by a flow of secondary ions with its own differential LET spectrum $\Phi_{\sec}(\Lambda)$ (here, $\Lambda$ is LET, for brevity). Then, the expected value of the error number per memory cell with a fixed critical charge $Q_C$ (or, the same, the soft error rate per bit (SER)) can be generally calculated as follows [15]

$$\mathrm{N}(Q_C)=\int \sigma(\Lambda \mid Q_C)\Phi_{\sec}(\Lambda)d\Lambda\,, \tag{2}$$

where $\sigma(\Lambda \mid Q_C)$ is the LET dependent direct ionization induced cross sections for secondary particles, $\Phi_{\sec}(\Lambda)$ is a differential LET spectrum of secondary particles for a given period of time. The spectrum $\Phi_{\sec}(\Lambda)$ is depends on the cross sections of the nuclear interactions, and also on the integrated circuit characteristics (layout, chemical composition, overlayer structure etc.) within the influencing volume with spatial scales $\sim L_R$ significantly exceeding the charge collection length.

The same value can be calculated as a result of the impact of primary nucleons (neutrons or protons) using the nucleon cross section concept $\sigma_n(\varepsilon_n \mid Q_C)$

$$\mathrm{N}(Q_C)=\int \sigma_n(\varepsilon_n \mid Q_C)\Phi_n(\varepsilon_n)d\varepsilon_n\,, \tag{3}$$

where $\Phi_n(\varepsilon_n)$ is the primary nucleon energy spectrum. Each energy of primary nucleon corresponds to its own LET spectrum of secondary particles which can be approximated somehow or simulated with GEANT4. The LET spectrum of secondary particles from nucleons of all energies can be written in the form

$$\Phi_{\sec}(\Lambda)=\int \alpha(\Lambda \mid \varepsilon_n)\Phi(\varepsilon_n)d\varepsilon_n, \tag{4}$$

where a conversion function $\alpha(\Lambda \mid \varepsilon_n)$ transform an energy spectrum of primary particles to a LET spectrum of secondary particles.

Equations (2) and (3) are fully consistent and equivalent under the linear integral relation

$$\sigma_n(\varepsilon_n \mid Q_C)=\int \alpha(\Lambda \mid \varepsilon_n)\sigma(\Lambda \mid Q_C)d\Lambda\,. \tag{5}$$

This can be verified directly by substituting (5) into (3) and changing the order of integration. An incident flux of primary nucleons with energy $\varepsilon_n$ produces the secondary ion flux with $\Phi_{\sec}(\Lambda)$. The LET spectrum of secondary particles can be considered as a good metrics when describing not only proton-induced but also neutron-induced SEUs. GEANT4 can be used as an appropriate tool to simulate the secondary particle LET spectrum $\Phi_{\sec}(\Lambda)$ at given proton or neutron energy spectra [16-18].

Each nucleon energy corresponds to its own form of the LET spectrum of secondary ions, which can be calculated using GEANT4. Such a procedure is very time consuming and therefore a simplified and flexible analytical model is highly desirable.

*2.2. Cross Section Modeling: Energy Dependence from LET Dependence*

For mono-energetic nucleons with a fluence $\Phi_n$ we have $\Phi(\varepsilon_n) = \Phi_n \delta(\varepsilon - \varepsilon_n)$ and $\Phi_{sec}(\Lambda) = \alpha(\Lambda | \varepsilon_n)\Phi_n$. Thus, the conversion function $\alpha(\Lambda | \varepsilon_n) = \Phi_{sec}(\Lambda)/\Phi_n$ is simply the LET spectrum of secondary particles normalized to a fluence of primary particles with a given energy. The conversion function can be written in a form

$$\alpha(\Lambda | \varepsilon_n) = \alpha_n(\varepsilon_n)\, p(\Lambda | \varepsilon_n), \tag{6}$$

where $p(\Lambda | \varepsilon_n)$ is a normalized to unity LET spectrum generated be primary nucleons with energy $\varepsilon_n$. The energy dependent efficacy $\alpha_n(\varepsilon_n)$ can be approximated as follows [19, 20]

$$\alpha_n(\varepsilon_n) \cong \Sigma(\varepsilon_n) N_{at} L_R(\varepsilon_n), \tag{7}$$

where $N_{at}$ is the atom density in material (~5×10$^{22}$ cm$^{-3}$ in silicon), $\Sigma(\varepsilon_n)$ is the total cross section of neutron-nuclear interactions, $L_R(\varepsilon_n)$ is a mean range of secondary particles. Strictly speaking, the secondary particle generation efficacy $\alpha_n(\varepsilon_n)$ should also include the average number of the products per a nuclear reaction, but this uncertainty can be absorbed by the uncertainty of the secondary particle range which can be considered as a fitting parameter. Assuming the maximum value $\Sigma(\varepsilon_n) \sim 10^{-24}$ cm$^2$, (see for example, [21]) and $L_R \geq 1$ μm, one gets approximately $\alpha_n \cong 10^{-5} - 10^{-6}$. This is an order of magnitude of a ratio of the maximal cross sections conditioned by nuclear reaction (typically, of order 10$^{-12}$-10$^{-14}$ cm$^2$ per bit) and heavy ion induced direct ionization (~ 10$^{-7}$ – 10$^{-9}$ cm$^2$).

Numerical Monte-Carlo, analytical and SRIM (see, e.g. [22-24]) simulations have shown that the LET of secondary particles are distributed approximately according to an exponential law up to a certain maximum value $\Lambda_{max}$ ~ 10-14 MeV-cm$^2$/mg. Our GEANT4 calculation demonstrate the same picture (see Fig. 2).

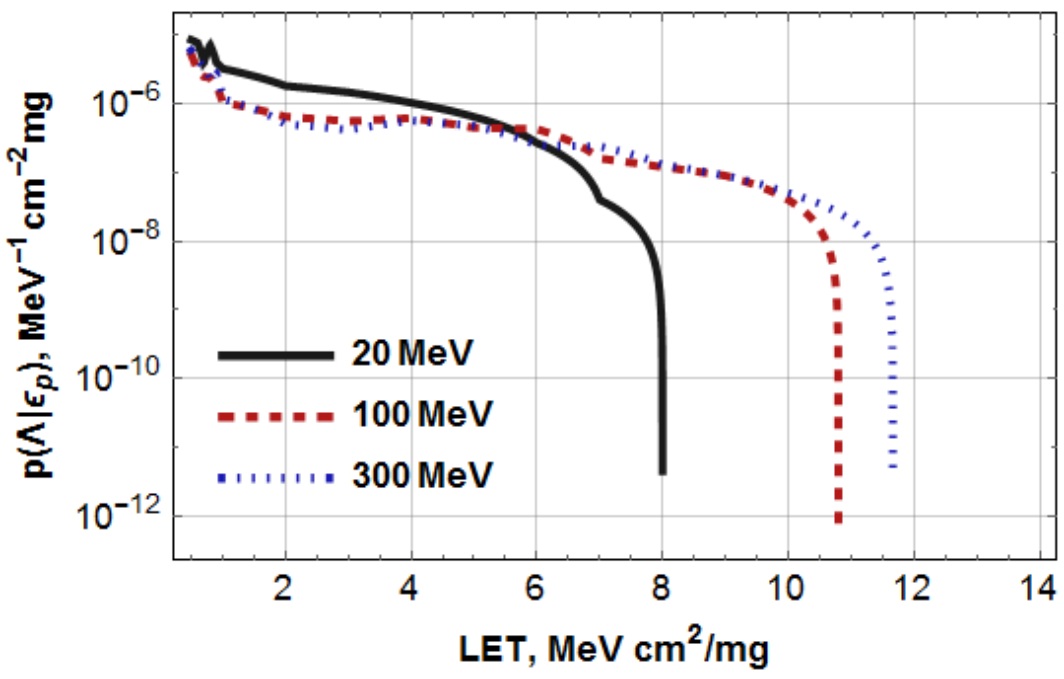

**Figure 2.** GEANT4 simulations of LET spectra of secondary particles at diffferent proton energies [14].

The LET spectra of secondary particles generally depend on many different factors: nucleon energies, IC layout and geometry, chemical composition of materials, etc. The uncertainty of these factors complicates and reduces the reliability of formally more accurate results. Therefore, we will focus in this work on simple analytical models with physically clear fitting parameters that can be refined using numerical calculations. The key approximation, which leads to a dramatic simplification of the simulation, is the effective LET approximation, in which the LET spectrum is approximated by a delta function $p(\Lambda | \varepsilon_n) \cong \delta(\Lambda - \Lambda_{eff}(\varepsilon_n))$, where $\Lambda_{eff}(\varepsilon_n)$ is an effective LET which is limited from above by some effective maximum value $\Lambda_{max}$~ 8-14 Mev-cm$^2$/mg. The introduction of a fitting parameter $\Lambda_{eff}(\varepsilon_n)$ allows us to avoid cumbersome numerical integration of ill-defined functions in (5). Indeed, the conversion function takes the form

$$\alpha(\Lambda | \varepsilon_n) \cong \Sigma(\varepsilon_n) N_{at} L_R(\varepsilon_n) \delta(\Lambda - \Lambda_{eff}(\varepsilon_n)), \tag{8}$$

and integration in (5) yields an explicit relation for nuclear induced SEU cross section through direct ionization cross section from secondary heavy ions.

$$\sigma_n(\varepsilon_n) \cong \Sigma(\varepsilon_n) N_{at} L_R(\varepsilon_n) \sigma\left(\Lambda_{eff}(\varepsilon_n) \mid Q_C\right) \quad (9)$$

The next paragraph will be devoted to the parametrization of SEU cross sections from direct ionization.

*2.3. Direct Ionization SEU Cross Section Parametrization*

With a decrease in the technological nodes, the threshold energy decreases and the single event mechanisms begins to play an ever-smaller role. Independence from the mechanism means an increase in the role of statistics, describing the processes without internal correlation. Particularly, the process of ionization and the transfer of energy from radiation to material becomes quantized and is determined more by general statistics. Based on statistical consideration we found that the SEU cross section (probability) per bit can be explicitly estimated by the value of the collected charge $\Delta Q$ during the passage of one ionizing particle [25]

$$\sigma(\Delta Q) = \frac{a_C}{\exp\left(\frac{Q_C}{\Delta Q}\right) - 1}, \quad (10)$$

where $a_C$ is the area of the memory cell. Generally, this a form of the geometrical, or Bose-Einstein distribution. In practice, the only thing that a researcher can experimentally set is the particle's LET at the input, which linearly related with $\Delta Q$. This makes it possible to rewrite (10) in an approximate form allowing direct comparison with test results

$$\sigma(\Lambda) = \frac{a_C}{\exp\left(\frac{2\Lambda_C}{\Lambda}\right) - 1} \cong \begin{cases} K_d(\Lambda - \Lambda_C), & \Lambda > \Lambda_C, \\ a_C e^{-\frac{2\Lambda_C}{\Lambda}}, & \Lambda < \Lambda_C, \end{cases} \quad (11)$$

where the slope and threshold of the quasi-linear part of the curve are directly measured experimental parameters connected with the circuit parameters through the relations

$$K_d = \frac{a_C}{2\Lambda_C}, \qquad \Lambda_C = \frac{\varepsilon_C}{\rho_{Si} t_{eff}}. \quad (12)$$

Here, the silicon mass density is $\rho_{Si} \cong 2.3$ g/cm$^3$, and $t_{eff}$ is the effective charge collection length. The "above-threshold case" $(\Lambda > \Lambda_C)$ in (11) describes the multiple cell upset cross sections $(\sigma \geq a_C)$ [26]. The "sub-threshold case" $\Lambda < \Lambda_C$ correspond the Hazucha-Svensson approximation [27]. It should be emphasized that hereinafter, the critical LET $\Lambda_C$ will always be understood as the result of an unambiguous interpolation of the linear part of the ion cross sections $\sigma(\Lambda)$ (not a Weibull curve parameter) related to $\varepsilon_C$ and $Q_C$ by (1) and (11) (see Fig. 3).

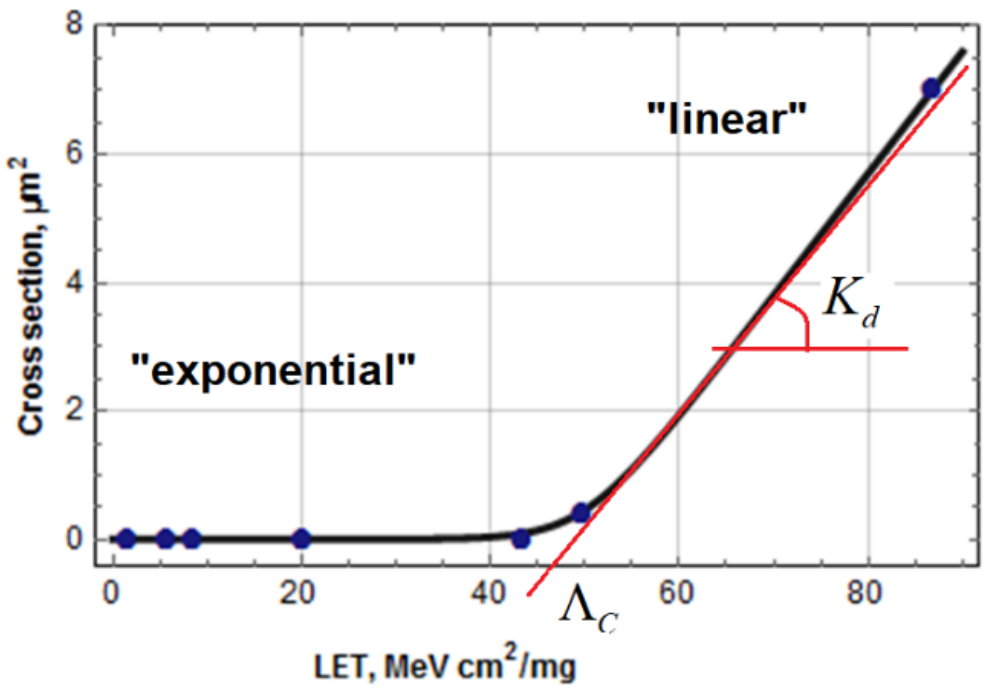

**Figure 3.** Typical SEU cross section vs LET dependence in linear scale [13].

**3. Results**

*3.1. Nucleon-Induced SEU Cross Sections*

The probability of soft errors is proportional to ionization energy deposition and $\Lambda_{eff}$ in the sensitive region. Experimental data and simulations indicate that zero nucleon energy corresponds to zero $\Lambda_{eff}$, and at high $\varepsilon_n$ the effective LET $\Lambda_{eff}$ will tend to its maximum value $\Lambda_{\max}$ ~ 8-14 MeV-cm²/mg (see Fig. 2). Therefore, it seems reasonable to parameterize $\Lambda_{eff}(\varepsilon_n)$ as a monotonically increasing function of the nucleon energy with saturation at $\Lambda_{\max}$, for example as follows

$$\Lambda_{eff}(\varepsilon_n) \cong \frac{1}{2}\left(\frac{1}{\Lambda_{\max}} + \frac{L_R \rho_{Si}}{\varepsilon_n}\right)^{-1} \cong \begin{cases} \dfrac{\varepsilon_n}{2\rho_{Si} L_R}, & \varepsilon_n < \Lambda_{\max}\rho_{Si} L_R, \\ \Lambda_{\max}/2, & \varepsilon_n > \Lambda_{\max}\rho_{Si} L_R, \end{cases} \tag{13}$$

where $L_R$ (of order of micrometers) is an effective range of secondary ions which can be used as a fitting parameter.

The average LET of secondary ions will increase only up to some threshold neutron energy $\varepsilon_n \le \varepsilon_{nT} = \Lambda_{\max}\rho L_R \cong 3.2(L_R/\mu m)$ MeV ≤ 10 MeV, and then saturates at the value $\Lambda_{\max}/2$. In fact, GEANT4 calculations show that the values $\Lambda_{\max}$ themselves are increasing functions of the energy of primary particles (see, Fig. 2) and we will consider $\Lambda_{\max}$ as a fitting parameter. Figure 4 shows rough approximations for the effective LET of secondary particle as functions of neutron energies.

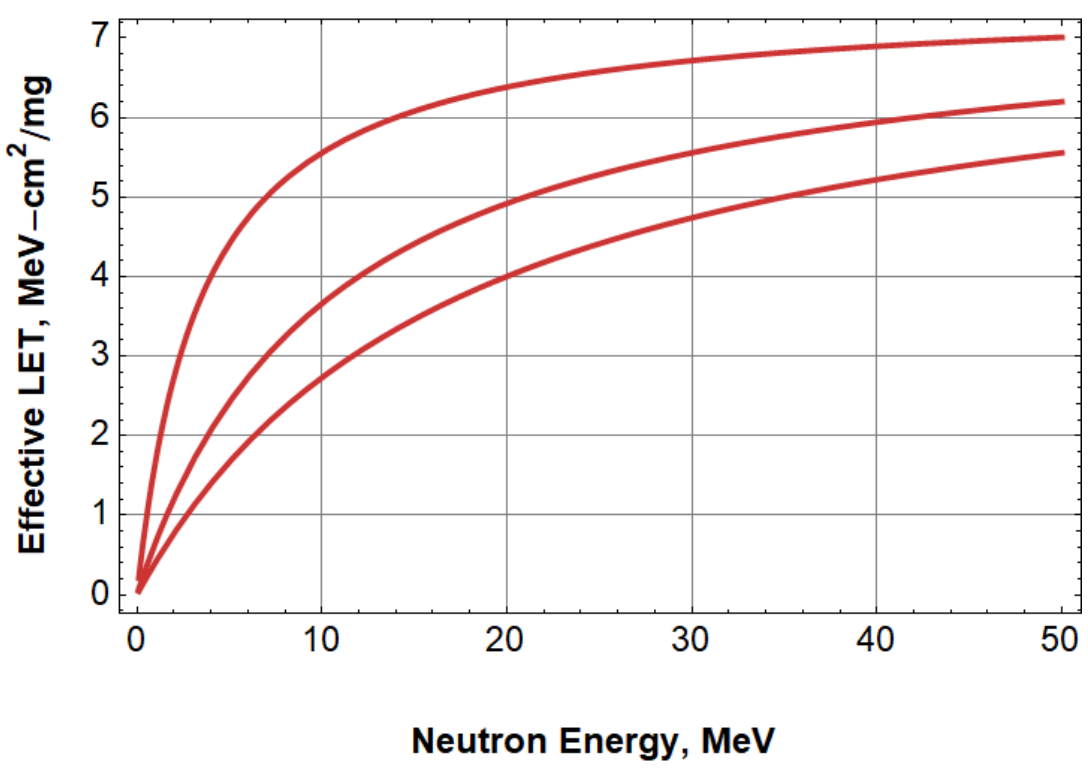

**Figure 4.** Approximate dependencies of the effective LET on neutron energy $\Lambda_{eff}(\varepsilon_n)$ simulated with (13) at $\Lambda_{max}$ = 15 MeV-cm²/mg for 3 ion ranges $L_R$ = 1, 3 and 5 μm.

Following (10) and (12), one can obtain an approximate expression for estimating the energy dependence of the neutron induced cross sections

$$\sigma_n(\varepsilon_n) = \frac{\alpha_n(\varepsilon_n) a_C}{\exp\left(2\Lambda_C/\Lambda_{eff}(\varepsilon_n)\right) - 1}, \tag{14}$$

where $\alpha_n$ can be estimated through (11). The neutron-induced nuclear reaction cross sections at relatively high energies ($\varepsilon_n \ge$ 10 MeV) are approximately constant $\Sigma_n^{\max}$ ~ 1-2 bn, whereas the low-energy ($\varepsilon_n \le$ 2 MeV) cross sections are usually much smaller, although the behavior of the curve can be very complex [28].

We will roughly approximate the energy dependence of the cross section for neutron nuclear interactions by the phenomenological formula

$$\Sigma(\varepsilon_n) \cong \frac{\Sigma_n^{\max}}{1 + \exp\left(-\dfrac{\varepsilon_n - \varepsilon_{n0}}{\delta\varepsilon_{n0}}\right)}, \tag{15}$$

where $\varepsilon_{n0}$ and $\delta\varepsilon_{n0}$ are fitting constants. Figures 5 and 6 show the neutron SEU cross sections as functions of $\varepsilon_n$ and circuit parameter $\Lambda_C$, analytically simulated with (13-15).

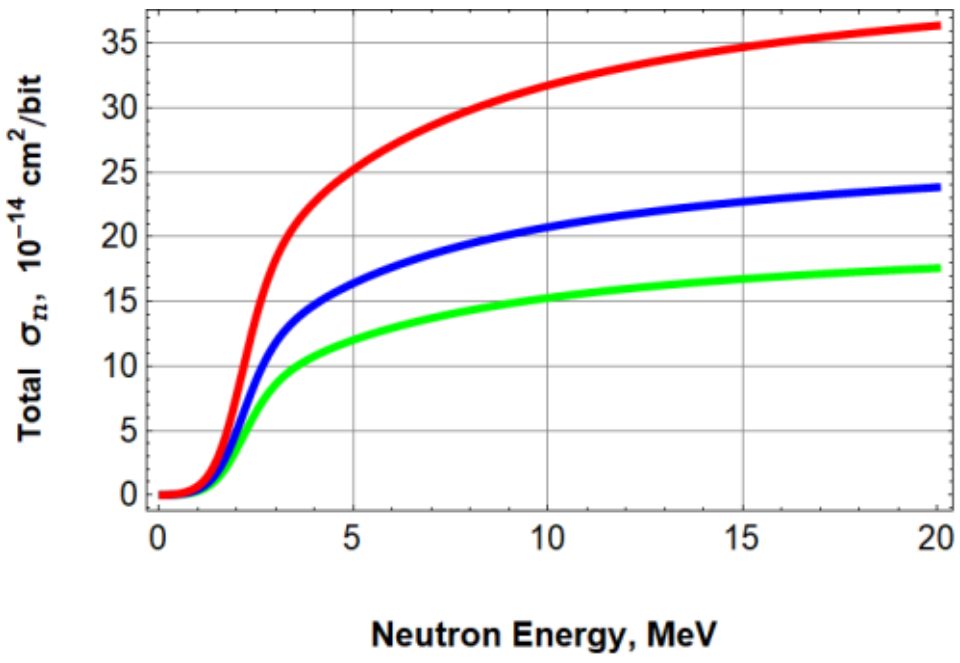

**Figure 5.** Energy dependencies of neutron-induced SEU cross sections per bit simulated for three values of critical LET (for $\Lambda_C$ = 0.2 (red), 0.3 (blue), 0.4(green) MeV-cm²/mg). Parameters: $a_C$ =0.5 µm², $\Lambda_{max}$ = 15 MeV-cm²/mg, $\varepsilon_{n0}$ = 2 MeV, $\delta\varepsilon_{n0}$ = 0.4 MeV.

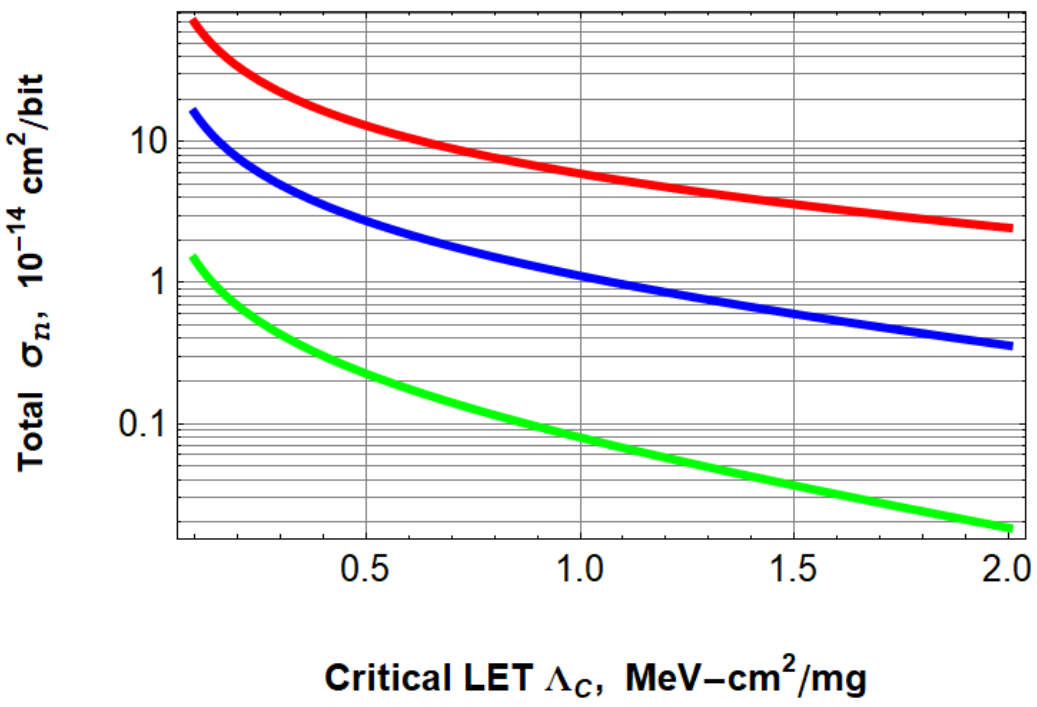

**Figure 6.** Neutron-induced SEU cross sections per bit simulated as functions of critical LETs for three values of neutron energy (for $\varepsilon_n$ = 1(green), 2(blue), 14 (red) MeV) with the same parameters.

The same equations (13-15) were used to validate the model using the experimental data for proton cross sections reported in [29] (see Fig. 7).

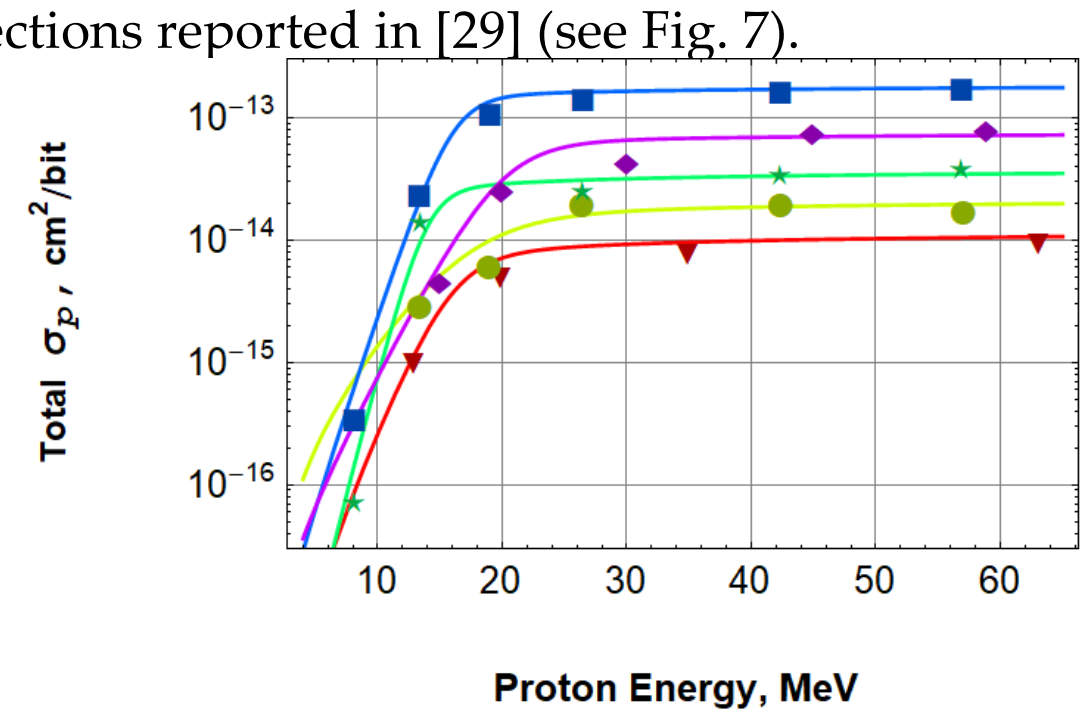

**Figure 7.** Comparison of experimental (points) and simulated (lines) data for proton-induced SEU cross sections (Lambert et al., 2009 [29]).

The data were simulated with fixed constants $a_C$ =1 µm², $\Lambda_{\max}$ = 14 MeV-cm²/mg, $L_R$ = 1 µm, $\Sigma_n^{\max}$ = 2×10$^{-24}$ cm², the logarithmic slope and the saturation onset position of the proton cross section curves in Fig. 7 were adjusted by a small change in the parameters $\delta\varepsilon_{n0}$ (from 1.5 to 3.5 MeV) and $\varepsilon_{n0}$ (from 14 to 20 MeV ). The maximum cross sections were adjusted by the value of $\Lambda_C$ (from 1.5 to 7.8 MeV-cm²/mg). Any refinement of input parameters (e.g. $a_C$ or $\Sigma_{n(p)}^{\max}$ ) will reduce the uncertainty of other fitting constants. Thus, unlike the traditional Bendel's and Weibull's empirical interpolation [30], the relation (14) is entirely based on transparent physical and circuit approximations and pa-rameters and therefore can be easily modified and recalculated for different IC nodes, chemical composition of the environment, and radiation characteristics.

### *3.2. Proton-Induced SEU Cross Sections*

The total cross section for proton-induced as a function of proton energy can be written as the sum of two cross sections due to nuclear reactions and direct ionization

$$\sigma_p \equiv \sigma_{p,nucl} + \sigma_{p,dir} = \\ = \frac{\alpha_p a_C}{e^{2\Lambda_C/\Lambda_{eff}(\varepsilon_p)} - 1} + \frac{a_C}{e^{2\Lambda_C/\Lambda_p(\varepsilon_p)} - 1} \quad (16)$$

where $\alpha_p(\varepsilon_p)$ is the secondary particle generation efficacy which can be simulated with GEANT4 for every specific configuration (see Fig. 8 from [14]) The proton efficacy can be parameterized in the same way as for neutrons $\alpha_p(\varepsilon_p) = N_{at}\Sigma(\varepsilon_p)L_R(\varepsilon_p)$.

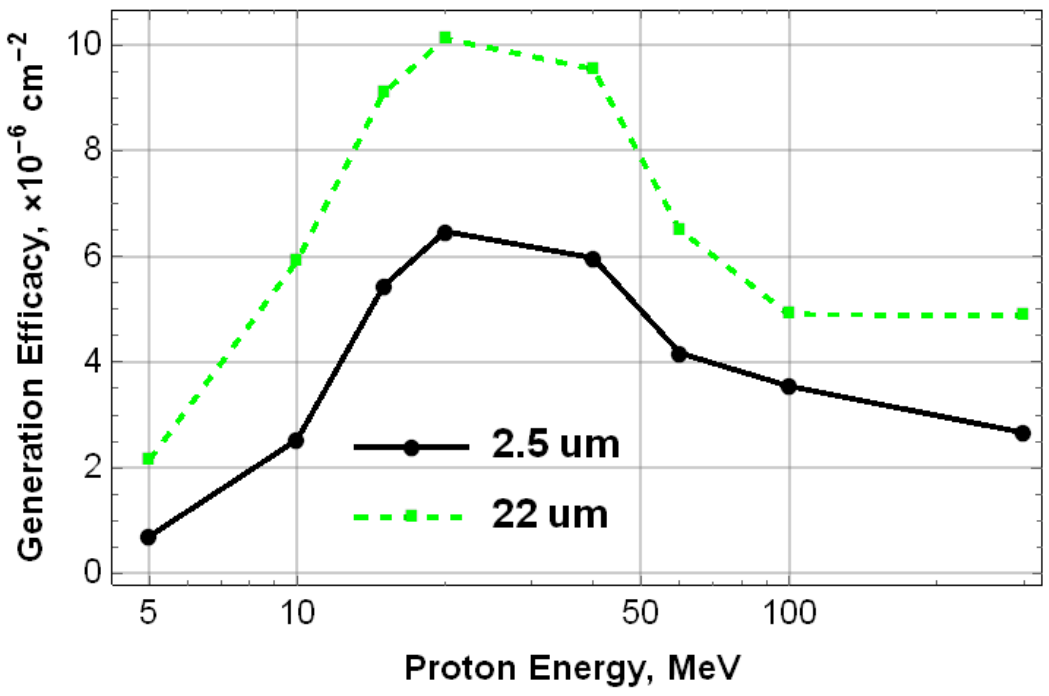

**Figure 8.** The efficacy of secondary particle generation $\alpha(\varepsilon_p)$ simulated with GEANT4 as functions of proton energies $\varepsilon_p$ for overlayer thickness 2.5 μm (solid line) and 22 μm (dashed line).

Then, the efficacy decrease at large proton energies can be explained by the energy dependence of the elastic Rutherford scattering $\Sigma_{Rutherford}(\varepsilon_p) \propto 1/\varepsilon_p$, which is the dominant mechanism of nuclear interactions at moderate proton energies. This explains the slight decrease in the SEU cross section observed at high proton energies. The efficacy for low-energy protons grows due to an increase in the range of secondary particles $L_R(\varepsilon_p)$. The low energy nuclear reaction cross sections for protons at low energies (< 10 MeV) are typically much smaller than for neutrons due to the presence of the Coulomb barrier. The detailed behavior of nuclear reactions induced SEU's cross sections for low energy protons is masked by the direct ionization induced SEUs. We will use for protons the same approximation for the effective LET of secondary particles as for neutrons at low energies. The results for the proton SEU's cross sections at high energies usually do not differ much from the case of neutron-induced upsets [31].

The LET $\Lambda_p(\varepsilon_p)$ of low energy protons can be simulated using SRIM. A characteristic feature of the proton LET is the presence of a sharp peak at $\varepsilon_p$ ~ 50 keV (see Fig. 9).

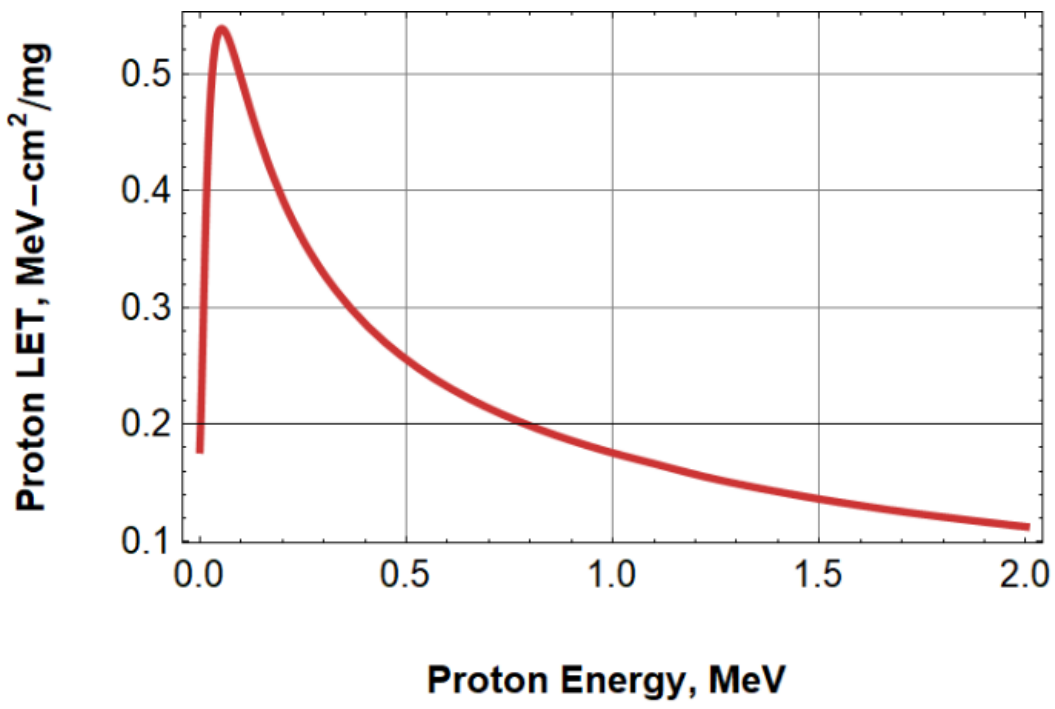

**Figure 9.** SRIM calculated dependence of LET as a function of proton energy.

For this reason, the calculated dependence of the cross section on direct ionization also must have a peak approximately at this energy. The SEU cross sections are monotonically decreasing functions of the parameter $\Lambda_C$ (see Fig. 10).

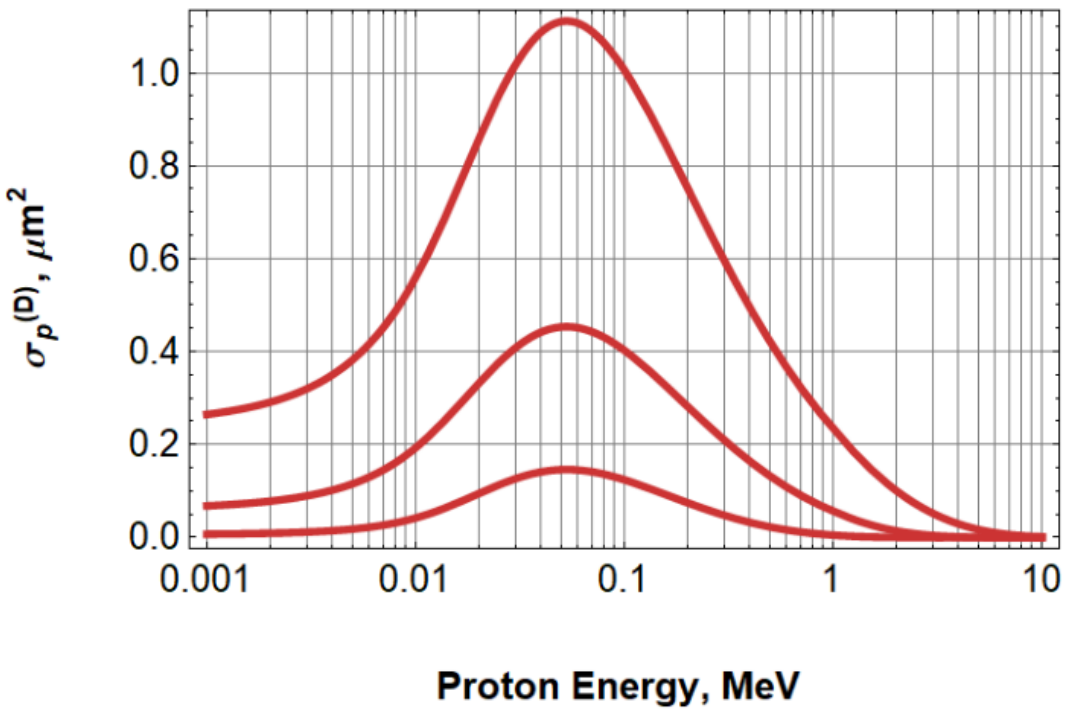

**Figure 10.** Non-monotonic view of direct ionization SEU cross sections for low-energy protons simulated with Eq. 16 for different Λc; $\Lambda_p(\varepsilon_p)$ was simulated using SRIM as in Fig. 9.

The specific form of the dependence of the total cross section on the proton energy, including direct and indirect ionization, is determined by the circuit parameter Λc, as shown in the illustrative simulation in Fig. 11.

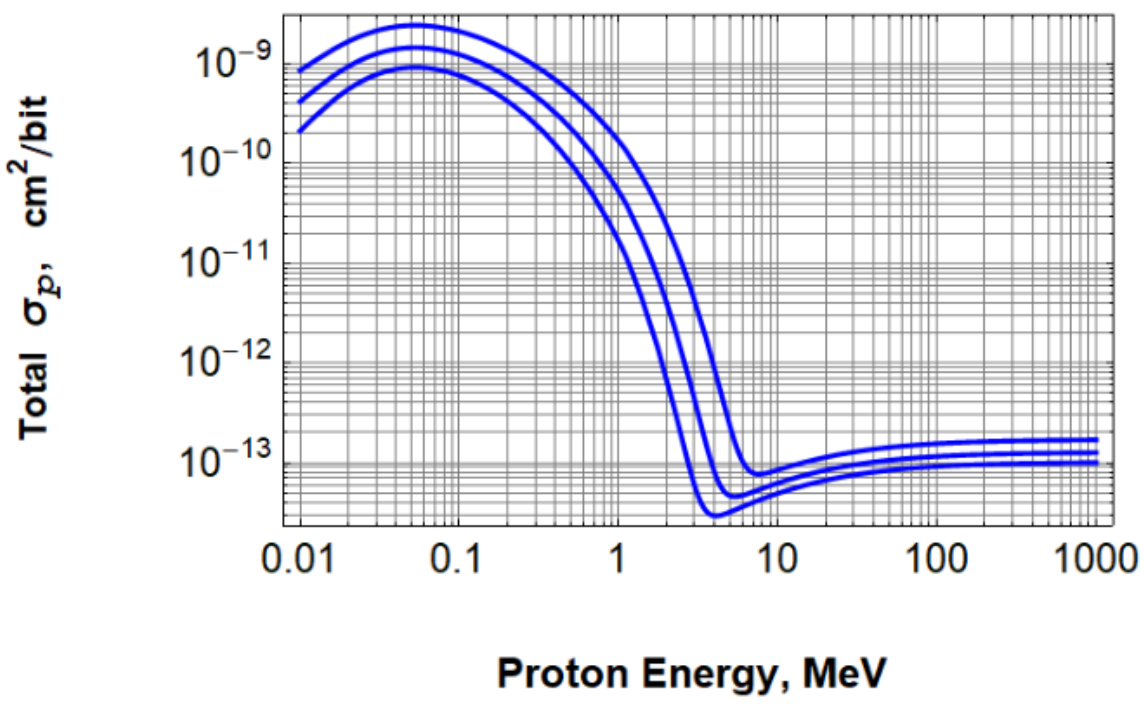

**Figure 11.** Typical proton-induced SEU's cross section as functions of energy, suimulated with Eq.16 for critical LETs Λc = 0.3, 0.4, 0.5 MeV-cm$^2$/mg (top down).

In practice, there is a problem associated with the fact that the low-energy proton energy spectra are formed by the specifics of local shielding, which makes it difficult simulations due to the difference in energies of external protons and directly ionizing internal protons. As the proton energy increases to about 10 MeV, the proton LET decreases and the total SEU cross section drops exponentially (see Eq. 16) to the level of nuclear-induced effects. The results of the model validation are shown in Fig. 12 where the comparison of calculation results and recent experimental data (adopted from Fig.1 in [32]) is presented.

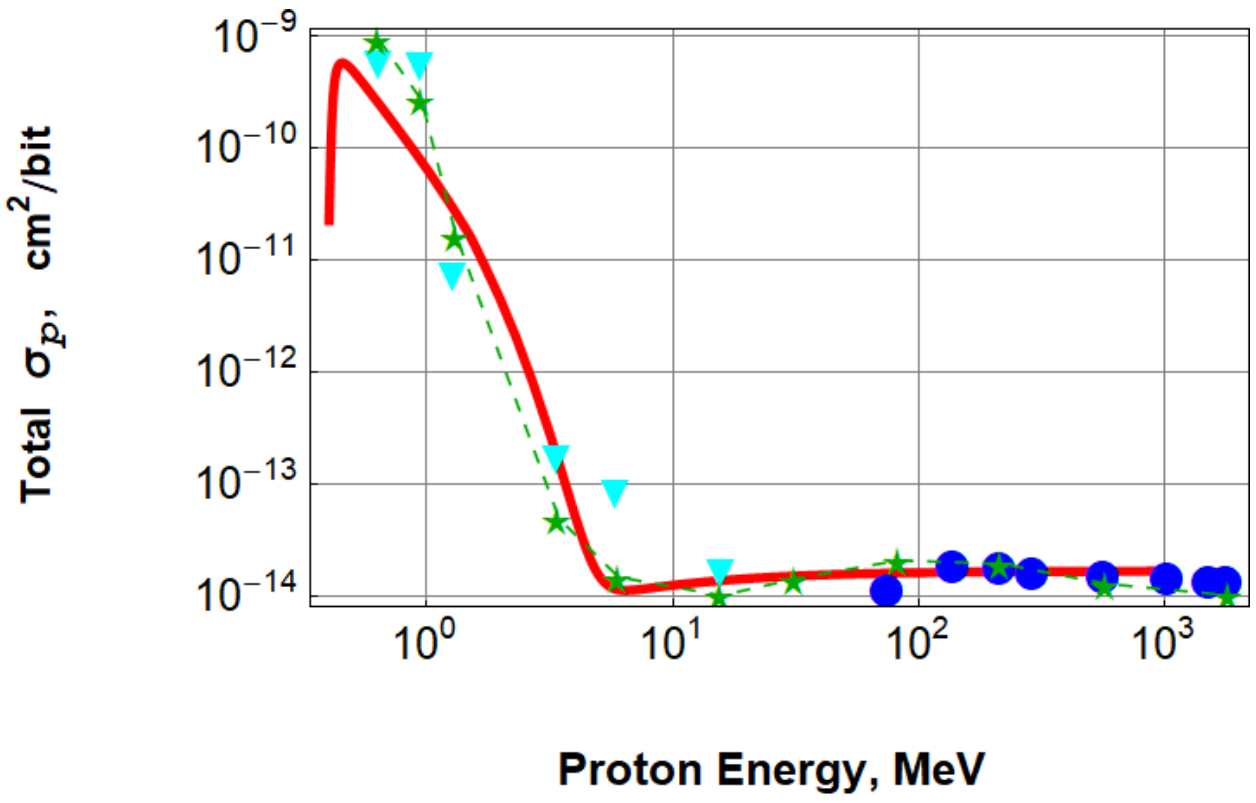

**Figure 12.** Comparison of experimental total proton cross sections vs energy (different points) and simulation with (16) (red solid line). Fitting

Simulated proton-induced SEU cross sections are in reasonable agreement in exponentially wide ranges of cross sections at low-, intermediate- and high-energy proton and 14 MeV neutron. Fitting parameters were $a_C$ =0.1 µm², $\Lambda_{\max}$ = 6 MeV-cm²/mg, $\Lambda_C$ = 0.4 MeV-cm²/mg, $L_R$ = 1 µm, $\Sigma_n^{\max}$ = 1×10$^{-24}$ cm².

## 4. Discussion

In this work, we proposed a general computational scheme for estimating the cross sections of SEU induced by direct ionization and nuclear interactions (Eqs. 2-5 and 10-11) and its simplified analytical implementation (Eqs. 8,9 and 13,15). The weakest point of the proposed simplified computational scheme is the method for roughly estimating the effective LET (Eq.13). A more rigorous approach involves averaging over the exponential LET spectrum of secondary particles (see Eqs. 5-6), which strictly speaking can only be calculated using the GEANT4 simulation. Any advanced numerical simulation makes sense in the presence of very detailed input data, which is not available in practice. Therefore, such calculations require cumbersome additional calibration, which is not much different from the simple fitting of one parameter within a simplified analytical model.

The main conceptual differences of the proposed approach compared to existing ones are as follows.

- Our proposed method consistently derives the nuclear-induced SEU cross section as a result of direct ionization by secondary particles. This allows the two failure mechanisms to be described in a unified manner within the same mathematical formalism.
- The generalized relation (10) is used instead of the Hazucha-Svensson formula, which is a special case of (10) for large values of the critical charge.
- This allows us to describe multiple cell upsets and the SEUs at very low energy deposits during direct ionization of low-energy protons.
- Unlike our approach, the traditional Burst Generation Rate (BGR) method proposed by Ziegler and Lanford [33] is entirely based on computation of nuclear effects and does not include critically important technological parameters such as a memory cell area $a_C$. The BGR method is valid only when the sensitive volume is so large that energy deposition can be taken to be local.

A new physics-based compact model, which makes it possible to simulate in a unified way the neutron and proton of cosmic ray induced SEU cross sections, including effects from nuclear reaction products and from direct ionization by low-energy protons, has been proposed and validated. To summarize, we list the main advantages of the described unified approach:

Simplicity and realism (analytical equations with physical and circuit parameters);

- Universality (both for neutrons and protons);
- Generality (the same approach to direct ionization and nuclear reaction induced effects);
- Compatibility with circuit simulations (critical charge/energy, cell area, etc.);
- Compatibility with radiation simulations (GEANT4, SRIM, etc.);
- Flexibility (can be refined by numerical simulations and adapted to suit different purposes.

**Author Contributions:** Conceptualization, Gennady Zebrev; Data curation, Vladimir Emeliyanov; Formal analysis, Rustem Useinov; Funding acquisition, Nikolay Samotaev; Investigation, Rustem Useinov; Methodology, Gennady Zebrev; Software, Artur Galimov and Artyom Sharapov; Supervision, Nikolay Samotaev and Alexander Rodin; Visualization, Dmitri Kazyakin; Writing – original draft, Gennady Zebrev and Rustem Useinov; Writing – review & editing, Nikolay Samotaev and Alexander Rodin.

**Funding:** This work was carried out within the framework of the state task of the Ministry of Science and Higher Education of the Russian Federation (subject No. FSWU-2022-0022 "Low-temperature ceramic technologies (LTCC) in microelectronics").

**Institutional Review Board Statement:** Not applicable.

**Informed Consent Statement:** Not applicable.

**Data Availability Statement:** Data sharing not applicable.

**Conflicts of Interest:** The authors declare no conflict of interest. The funders had no role in the design of the study; in the collection, analyses, or interpretation of data; in the writing of the manuscript; or in the decision to publish the results.